\documentclass[twocolumn,aps,prl,groupedaddress]{revtex4}
\usepackage{epsfig,amssymb,amsmath}
\usepackage[hypertex,linkcolor=red]{hyperref}

\usepackage{color}

\def\comment#1{}
\def\labell#1{\label{#1}}
\def\vacuum{\emptyset}
\def\section#1{{\par\em #1:--- }}
\def\>{\rangle}\def\<{\langle}
\def\togli#1{}

\begin{document}
\title{Quantum metrology: why entanglement?}  \author{Lorenzo Maccone}
\affiliation{ \vbox{ Dip.~Fisica ``A.~Volta'', Univ.~of Pavia, via
    Bassi 6, I-27100 Pavia, Italy}} 
\begin{abstract}
  We show why and when entanglement is needed for quantum-enhanced
  precision measurements, and which type of entanglement is useful. We
  give a simple, intuitive construction that shows how entanglement
  transforms parallel estimation strategies into sequential ones of
  same precision. We employ this argument to generalize conventional
  quantum metrology, to identify a class of noise whose effects can be
  easily managed, and to treat the case of indistinguishable probes
  (such as interferometry with light).
\end{abstract}
\maketitle

{ To measure a parameter $\varphi$ that identifies some physical
  transformation $U_\varphi$, we can sample the transformation with a
  probe system and then measure the change in the state of the probe.
  To reduce the statistical error, we repeat the sampling $N$ times
  and average the results. Quantum metrology encompasses a large class
  of estimation procedures where quantum effects such as entanglement
  and squeezing are used to enhance the precision of such measurement
  over what would be possible with classical resources
  \cite{science,natphot}. Such techniques are by now well established
  \cite{caves1,caves2,qmetr,wiseman,caves3,rosetta,caves4,wineland}
  and have produced impressive results
  \cite{boixo,gross,geremia,pryde,luiz,walmsley,chiara,guta,kitagawa,shaji,mitchell,hayashi}.
  Yet, the reason {\em why} entanglement is necessary has not been
  settled.  Here we present a simple, intuitive argument on why and
  when entanglement is necessary and, additionally, use it to derive
  some new results in quantum metrology.} Our argument shows how
entanglement can convert a parallel-sampling strategy (where $N$
probes sample the system in parallel) to a sequential strategy, which
is known to achieve high precision, and viceversa.  While the
equivalence of parallel and sequential strategy was already shown in
Ref.~\cite{boixo} using the theory of tensor networks, our
construction is much simpler, as it is based on a trivial algebraic
identity: this simplicity is key to give an intuition on what quantum
metrology means, to obtain some new results, and to reobtain more
easily some known ones: e.g., we show how one can easily generalize
the conventional quantum metrology framework; which kinds of errors
are simple to manage; how the theory accommodates indistinguishable
probes (e.g.~bosons for light interferometry); which types of
entanglement are useful; how non-asymptotic results arise in quantum
metrology; etc.

In the framework of quantum metrology we are given a ``black box''
that performs a unitary operation $U_\varphi=e^{i\varphi H}$, where
$\varphi$ is the parameter to be estimated and $H$ is its generator of
translations. What is the best precision with which we can determine
$\varphi$?  Call $|0\>$ and $|1\>$ the states that correspond to the
minimum and maximum eigenvalues of $H$. The existence of such states
is guaranteed only in finite-dimensional Hilbert spaces, and we will
restrict to this case here (see Refs.~\cite{heisbound,heisboundzz} for
some infinite-dimensional results).  If we sample the system
sequentially $N$ times with a single probe (Fig.~\ref{f:qmet}a), we
can perform a measurement of $\varphi$ as follows: (a)~prepare the
probe in the state $|+\>$ or $|-\>$, $|\pm\>\equiv |0\>\pm|1\>$ (we
will neglect state normalizations); (b)~evolve it using a sequence of
$N$ unitaries as $U^N_\varphi|\pm\>=|0\>\pm e^{iN\varphi}|1\>$;
(c)~project the final state onto the initial one: the probability that
they coincide is $p(\varphi)=\cos(N\varphi/2)$ (Ramsey interferometry
corresponds to the case $N=1$ \cite{rosetta}). By repeating the above
procedure $\nu$ times, the experimental probability $p(\varphi)$,
whence $\varphi$, are inferred with an error that scales as
$1/(N\sqrt{\nu})$ \cite{luis,qmetr} where the factor $\sqrt{\nu}$
comes from the central limit theorem and the factor $N$ comes from the
fact that the sequential procedure is equivalent to measuring the
parameter $N\varphi$: an error $\Delta$ on $N\varphi$ corresponds to
an error $\Delta/N$ on $\varphi$.\togli{ [It is important that the
  procedure works equally well if one starts with $|+\>$ or $|-\>$.]}
In contrast, if we prepare the $N\nu$ probes in the state $|+\>$ or
$|-\>$, use them to sample the system in parallel and average the
results (Fig.~\ref{f:qmet}b), from the central limit theorem if
follows that we achieve an error that asymptotically scales as
$1/\sqrt{N\nu}$-- the ``standard quantum limit'' (SQL).
Surprisingly, if the $N$ probes are prepared in an entangled state
(Fig.~\ref{f:qmet}c), the quantum Cramer-Rao bound can be used to show
that one can asymptotically achieve the same $1/N$ limit
\cite{qmetr,caves1,caves2} of the sequential strategy-- the
``Heisenberg bound''. A typical quantum metrology experiment:
(a)~prepare $N$ probes in the entangled state $|0\>^{N}+|1\>^{N}$;
(b)~evolve it by applying $U_\varphi$ to each probe in parallel:
$U^{\otimes N}(|0\>^{N}+|1\>^{N})=|0\>^{N}+e^{iN\varphi}|1\>^{N}$;
(c)~project on the initial state as above to estimate the parameter
$N\varphi$: as before, the probability is
$p(\varphi)=\cos(N\varphi/2)$, whence $\varphi$ can be estimated with
an error that scales as $1/(N\sqrt{\nu})$ as in the sequential case
above. While this is well known, the reason why entanglement helps was
not explored in depth: typically one would say that this is the effect
of the ``global'' nature of the entangled state that cannot be
factored into separate contributions for each probe. Here we give a
simple construction that shows how one can convert a
parallel-entangled strategy into a sequential one and viceversa,
proving that this is a reason why they achieve the same error scaling.
We show that classical correlation among the probes is insufficient.

\begin{figure}[hbt]
\begin{center}
\epsfxsize=.6\hsize\leavevmode\epsffile{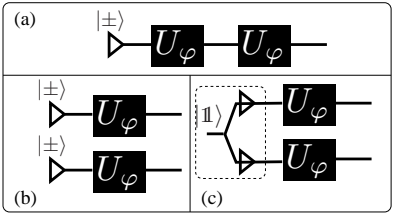}
\end{center}
\vspace{-.5cm}
\caption{Possible strategies for estimating the parameter $\varphi$
  that a unitary transformation $U_\varphi$ (black boxes) encodes into
  a probe system (triangles). (a)~Sequential strategies: a single
  probe samples the $N$ boxes sequentially, effectively measuring
  $N\varphi$ and achieving an error $\propto N^{-1}$ on the estimation
  of $\varphi$.  (b)~Classical (unentangled) parallel strategies: we
  use $N$ probes and average the measurement results: the central
  limit theorem tells us that the error on $\varphi$ scales as
  $N^{-1/2}$.  (c)~Entangled strategy (quantum metrology): the $N$
  probes are prepared in a joint entangled state, but sample the black
  boxes separately: the error scales as $N^{-1}$. }
\labell{f:qmet}\end{figure}

\section{The argument} We start detailing our construction for $N=2$
and then extend it to arbitrary $N$ by induction. Given an operator
$C=\sum_{ij} C_{ij}|i\>\<j|$ ($\{|i\>\}$ a basis), we can define a
state $|C\>\equiv\sum_{ij}C_{ij}|i\>|j\>$. It can be used
\cite{paolop} for the following identity\begin{eqnarray}
A\otimes B|C\>=
\textstyle{\sum_{ijkl}}A_{ki}B_{lj}C_{ij}|k\>|l\>=|ACB^T\>,
\labell{id}\;
\end{eqnarray}
where $^T$ indicates the transpose in the $|i\>$ basis. Consider the
quantum metrology protocol (Fig.~\ref{f:qmet}c) for $N=2$, and apply
this identity twice:
\begin{eqnarray}
(U_\varphi\otimes U_{\varphi'})|\openone\>=
|U_\varphi U_{\varphi'}^T\>=
(U_\varphi U_{\varphi'}^T\otimes\openone)|\openone\>
\labell{arg}\;,
\end{eqnarray}
where $|\openone\>=|00\>+|11\>$, and we consider the general case
where $\varphi$ and $\varphi'$ may differ.  Choose $|i\>$ as the
eigenbasis of $H$, so $U_\varphi=e^{i\varphi H}$ is diagonal, and
$U_\varphi^T=U_\varphi$.  Moreover, we can exploit the entanglement of
$|\openone\>$ to measure the second system in the $\{|+\>,|-\>\}$
basis (we consider only the subspace spanned by $|0\>$ and $|1\>$) and
project the first system in the same state, since
$|00\>+|11\>=|++\>+|--\>$. This means that after the measurement, the
evolution in Eq.~\eqref{arg} can be written as $U_\varphi
U_{\varphi'}|\pm\>$, where the initial state is $|+\>$ or $|-\>$ with
probability 1/2, depending on the measurement outcome. This argument
shows how the parallel entangled strategy $(U_\varphi\otimes
U_{\varphi'})|\openone\>$ can be reduced to the sequential one
$U_\varphi U_{\varphi'}|\pm\>$ for $N=2$, see Fig.~\ref{f:ff}.

\begin{figure}[hbt]
\begin{center}
\epsfxsize=1.\hsize\leavevmode\epsffile{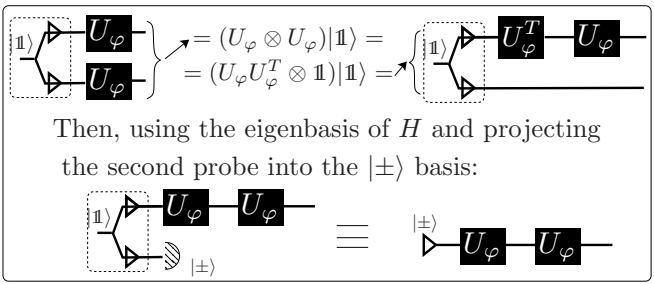}
\end{center}
\vspace{-.5cm}
\caption{Argument to transform a parallel entangled strategy into a
  sequential one for $N=2$, see Eq.~\eqref{arg}.
}\labell{f:ff}\end{figure}

Why is entanglement
necessary? We must require that perfect correlation is present in two
complementary basis, the $|0,1\>$ basis (to ensure that $U_\varphi$ is
diagonal) and the $|\pm\>$ basis (to ensure that the measurement on
one probe projects the other in an appropriate initial state). An
intuitive definition of entanglement is ``correlation on a property
that does not (cannot) yet exist'', namely, a perfect correlation on
complementary properties such as the two bases above.  In
fact, suppose we request that there is correlation only on the
$|0,1\>$ basis using the mixed state $|00\>\<00|+|11\>\<11|$), that
can be written as an equally weighted mixture of
$|\openone\>=|00\>+|11\>$ and $|\sigma_z\>=|00\>-|11\>$, each with
probability 1/2. We can use the construction of Eq.~\eqref{arg}
obtaining a final state of the two probes given by $(U_\varphi
U_\varphi\otimes\openone)|\openone\>$ or $(U_\varphi\sigma_z
U_\varphi\otimes\openone)|\openone\>$ each with probability 1/2. After
the projection of the second probe into the $|\pm\>$ basis, these
become an equally weighted mixture of $|0\>\pm e^{i2\varphi}|1\>$ and
$|0\>\mp e^{i2\varphi}|1\>$, namely the state $|0\>\<0|+|1\>\<1|$,
independent of $\varphi$. In contrast, suppose we request correlation
only on the $|\pm\>$ basis using the state
$|\!++\>\<++\!|+|\!--\>\<--\!|$, namely an equally weighted mixture of
$|\openone\>=|\!++\>+|\!--\>$ and $|\sigma_x\>=|\!++\>-|\!--\>$ (where
$\sigma_x=|0\>\<1|+|1\>\<0|$ in the $|0,1\>$ basis). Again using
Eq.~\eqref{arg}, this is equivalent to having a final state of the two
probes as $(U_\varphi U_\varphi\otimes\openone)|\openone\>$ or
$(U_\varphi\sigma_x U_\varphi\otimes\openone)|\openone\>$ each with
probability 1/2, which, after the projection, become an equally
weighted mixture of the four states $|0\>\pm e^{i2\varphi}|1\>$ and
$|0\>\pm|1\>$, again $|0\>\<0|+|1\>\<1|$ independent of $\varphi$.
[Here the SQL is recovered if one does not average over the
projective-measurement result: that procedure is equivalent to the
classical parallel strategy of Fig.~\ref{f:qmet}b.]

We have shown that for $N=2$ we can easily convert a parallel
entangled strategy into a sequential one, proving that the two have
the same estimation precision. We now extend this to arbitrary $N$ by
induction: suppose (induction hypothesis) that the parallel entangled
strategy $\bigotimes_{j=1}^NU_{\varphi_j}(|0\>^N+|1\>^N)$ has the same
estimation precision as the sequential strategy
$\prod_{j=1}^NU_{\varphi_j}|\pm\>$, we need to prove that the same
applies for $N+1$. We can write
\begin{eqnarray}
  &&{\textstyle\bigotimes_{j=1}^{N+1}}U_{\varphi_j}(|0\>^{N+1}+|1\>^{N+1})
  \nonumber\\&&
={\textstyle\bigotimes_{j=1}^{N-1}}U_{\varphi_j}\otimes \bar
  U_\theta(|0\>^{N-1}|\bar 0\>+|1\>^{N-1}|\bar 1\>)
\labell{induz}\;,
\end{eqnarray}
where $|\bar 0\>=|00\>$, $|\bar 1\>=|11\>$, and
\begin{eqnarray}
&&  \bar U_\theta\equiv U_{\varphi_N}\otimes U_{\varphi_{N+1}}=
e^{i\lambda}\left(\begin{matrix}1&0\cr
      0&e^{i\varphi_{N}}\end{matrix}\right)\otimes\left(\begin{matrix}1&0\cr
      0&e^{i\varphi_{N+1}}\end{matrix}\right)\nonumber\\&&
=e^{i\lambda}\left(\begin{matrix}1& & & \cr & e^{i\varphi_{N+1}}& & \cr
 & & e^{i\varphi_{N}} & \cr  & & &e^{i(\varphi_{N}+\varphi_{N+1})} 
\end{matrix}\right),
\labell{uu}\;
\end{eqnarray}
since $U_{\varphi_j}$'s are diagonal in the $|0,1\>$ basis.  It follows
that the subspace spanned by $|\bar 0\>=|00\>$ and $|\bar 1\>=|11\>$
is invariant for application of $\bar U_\theta$, which, when
restricted to such subspace, can be expressed as the $2\times 2$
matrix
\begin{eqnarray}
U_\theta=e^{i\lambda}\left(\begin{matrix}1&0\cr
    0&e^{i(\varphi_{N}+\varphi_{N+1})}\end{matrix}\right)=
U_{\varphi_N}U_{\varphi_{N+1}}
\labell{ut}\;.
\end{eqnarray}
Restricting ourselves to such invariant subspace, we can use the
induction hypothesis to conclude that the last line of \eqref{induz}
will have the same estimation precision as
$\prod_{j=1}^{N-1}U_{\varphi_j}U_\theta
|\pm\>=\prod_{j=1}^{N+1}U_{\varphi_j} |\pm\>$, which proves the
inductive step, see Fig.~\ref{f:ffb}. [Note that in this proof it is
not necessary to require that the matrix $\bar U_\theta$ is diagonal,
but only that it is invariant for the $|\bar 0\>$, $|\bar 1\>$
subspace: an anti-diagonal tensor product would work.]
\begin{figure}[hbt]
\begin{center}
\epsfxsize=.95\hsize\leavevmode\epsffile{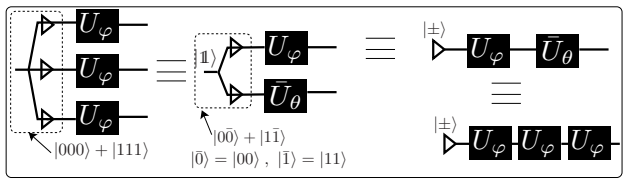}
\end{center}
\vspace{-.5cm}
\caption{Extension of the argument of Fig.~\ref{f:ff} to arbitrary
  $N$.  }\labell{f:ffb}\end{figure}

\section{Consequences} We now explore some of the consequences that
can be derived from our argument.

(a)~Useful entanglement: Up to now we have used as entangled initial
state the GHZ, $(|0\>^N+|1\>^N)$, as it optimizes the quantum
Cramer-Rao bound. Our construction shows that entanglement of the type
$|\Psi_N\>=|0\>^N+e^{i\lambda}|1\>^N$ (with real $\lambda$) is the
only useful one (with the appropriate amendments when probes are
indistinguishable, see below).  In fact, suppose we use an arbitrary
entangled state $|E\>$ in Eq.~\eqref{arg}, it becomes $(e^{i\varphi
  H}\otimes e^{i\varphi H})|E\>=(e^{i\varphi H} E e^{i\varphi
  H}\otimes\openone)|\openone\>$, where we want the right hand term to
correspond to a sequential strategy $e^{2i\varphi H}(|0\>\pm
e^{i\lambda}|1\>)$ (the sequential strategy works iff the initial
state is of the form $|0\>\pm e^{i\lambda}|1\>$ even though for
simplicity we considered only the cases $|\pm\>$ above).  Namely, we
require that $e^{i\varphi H} E e^{i\varphi H}|\pm\>=e^{2i\varphi
  H}(|0\>\pm e^{i\lambda}|1\>)$, which implies that $E=|0\>\<0|+
e^{i\lambda}|1\>\<1|$, since $H$ is diagonal in the $|0,1\>$ basis.
Namely, we need a state $|E\>=|\Psi_2\>$. This argument, and
the fact that achieving the quantum Cramer-Rao bound requires a
superposition of global maximum and minimum eigenstates \cite{qmetr},
shows that quantum metrology {\em requires} entanglement of the form
$|\Psi_N\>$.

(b)~Noise resilience: using our framework, we can detail a class of
noise that is simple to handle (in general, the treatment of quantum
metrology in the presence of noise is highly nontrivial
\cite{luiz,chiara,rafal,optdowlin,durkin,walmsley}). Consider the
entangled parallel strategy for $N=2$ and suppose that the two probes
are subject to arbitrary noise maps described by Kraus operators
$A_k$ and $B_k$ respectively. We can still convert a parallel strategy
into a sequential one along the lines of Eq.~\eqref{arg} as
\begin{eqnarray}
&&  \sum_{jk}(A_k\otimes B_j)|\openone\>
\<\openone|(A_k^\dag\otimes B_j^\dag)=
  \sum_{jk}|A_k B_j^T\>\<A_k^* B_j^\dag|\nonumber\\&&
=
  \sum_{jk}(A_k B_j^T\otimes\openone)
|\openone\>\<\openone|(B_j^*A_k^\dag\otimes\openone)
\labell{noise}\;,
\end{eqnarray}
i.e.~a noise map only on the first probe if
$\sum_jB_jB_j^\dag=\openone$ (a unital map, as when
$B_j\propto\sigma_\alpha$, the Pauli matrices), since then
$\sum_{jk}B_j^*A_k^\dag A_kB_j^T=\openone$, with
$B^*\equiv(B^T)^\dag$. So the first step of the induction can be
applied also to unital maps in addition to the unitaries we discussed
above. However, the extension to $N>2$ fails in general, as there is
typically no invariant subspace to construct a reduced map as the one
in \eqref{ut}: one must require that $A_k\otimes B_j$ is a diagonal or
an anti-diagonal matrix for all $k,j$. An example of the latter is the
bit flip with phase noise where the Kraus operators are
$A_0\propto\sigma_x$ and $A_1\propto\sigma_y$.  An example of the
former (in addition to the noiseless case) is phase flip noise
(dephasing) with Kraus operators $A_0\propto\openone$ and
$A_1\propto\sigma_z$, which was already identified in \cite{boixo} as
a noise model whose parallel and sequential performance match. It was
known for a long time \cite{chiara} that an arbitrary small amount of
dephasing would ruin any advantage for frequency measurement
\cite{luiz,guta,kitagawa,shaji} whereas the same identical noise does
not affect phase estimation as much \cite{natphot,shaji}.  Our
construction explains this curious performance: since the entangled
parallel strategy is (for this noise model) equivalent to the
sequential one, it means that the entangled strategy's performance is
equivalent to one that takes $N$ time as long: namely, its noise
sensitivity is $N$ times faster \cite{chiara}.  While this is not a
problem for estimating a phase (which can be done in arbitrarily short
time), it is a problem for estimating frequency which requires a
measurement duration that voids any advantage from entanglement
\cite{natphot,chiara,shaji}.  The calculation of the quantum
Cramer-Rao bound for frequency-estimation on a single qubit subject to
dephasing (the optimal sequential strategy) is given in
Ref.~\cite{shaji}, whence one can see it matches with the optimal
entangled strategy (although that was not observed there).\comment{In
  fact, Eq.~(4.13) of \cite{shaji} gives the optimal sequential
  strategy. In the case of one qubit it gives a lower bound to the
  determination of frequency $\delta g\geqslant
  e^{\gamma_2T}/T\sqrt{\nu}$ (choosing $n=1$).  Since this is the
  sequential strategy, we must multiply both the decay rate $\gamma_2$
  and the frequency $g$ by $n$, thus we obtain $n\delta g\geqslant
  e^{n\gamma_2T}/{T}\sqrt{\nu}$. This matches the Cramer-Rao bound for
  the parallel entangled strategy, given in Eq.~(4.30), namely $\delta
  g\geqslant e^{n\gamma_2T}/Tn\sqrt{\nu}$, as expected. However, note
  that in \cite{shaji} a more general noise model was analyzed, where
  not only dephasing, but also decay is present. However, to make the
  global map CP, one needs to choose ${T}_2\leqslant 2T_1$,
  i.e.~$\gamma_1\geqslant 2\gamma_2$, see below Eq.~(4.3). This means
  that dephasing ${T}_2$ time is shorter and dominates the evolution,
  so that $\gamma_2$ is the only noise parameter appearing in the two
  above bounds.}  Our construction thus identifies some noise
models whose effect is equivalent for the sequential and parallel
strategies: the unital maps (in the $|0\>$, $|1\>$ invariant subspace)
where all Kraus operators are either diagonal or anti-diagonal, as the
examples above.  It is then simple using our construction to find the
states that optimize the precision in the presence of these classes of
noise.

(c)~Indistinguishable probes: the above discussion focused on the case
of distinguishable probes, where a tensor product structure (TPS) is
clear: each probe is a system with its own Hilbert space. If the
probes are indistinguishable particles (e.g.~in optical
interferometry, where typically one considers single-photon probes),
such structure is absent, and one typically resorts to the Fock space
that enforces the symmetry properties required (symmetry for exchange
of Bosons and anti-symmetry for exchange of Fermions). In the absence
of a TPS, the very concept of entanglement is ill defined
\cite{paolo,paoloseth}, and it is known that entanglement is not
necessary to beat the SQL in interferometry. For example, the
unentangled ``N0'' state $|N\>+|\vacuum\>$, where $|N\>$ is a Fock
state of $N$ photons and $|\vacuum\>$ is the vacuum state (we restrict
ourselves to a finite-dimensional subspace of the radiation Hilbert
space) can achieve a phase sensitivity beyond the SQL: its evolution
through $e^{i\varphi a^\dag a}$ ($a$ being the annihilation operator
of the mode) yields $|\vacuum\>+e^{iN\varphi}|N\>$, as in the
sequential or entangled strategies of Figs.~\ref{f:qmet}a,c.  Because
such a state cannot be attained in the presence of super-selection
rules \cite{rudolph} and it only samples an absolute phase, in
quantum-enhanced interferometry one typically considers the two mode
``N00N'' state \cite{rosetta} $|N,\vacuum\>+|\vacuum,N\>$, which
samples a relative phase between the two modes $a,b$ of the
interferometer through the evolution $e^{i\varphi (a^\dag a-b^\dag
  b)}$. While clearly $|N,\vacuum\>+|\vacuum,N\>$ has mode
entanglement, the state $|N\>+|\vacuum\>$ has no entanglement. Yet,
they can both be easily cast in the framework we have used up to now,
by recalling that for Boson probes, one needs to consider only the
subspace of the global Hilbert space that is symmetric for exchange of
the probes, which can be done by moving to the Fock space
\cite{teller}: in each mode we have that the tensor product of minimum
and maximum eigenstates $|0\>^N$ and $|1\>^N$ is symmetrized as
$|\vacuum\>$ and $|N\>$, respectively. Analogously, in the two mode
case, the symmetrized minimum and maximum eigenstates for the two-mode
Hamiltonian $a^\dag a-b^\dag b$ restricted to the $N$-photon subspace
is $|\vacuum,N\>$ and $|N,\vacuum\>$, respectively (the symmetrization
must be performed in each mode separately, since photons in different
modes are distinguishable). It is hence clear that N00N state
interferometry falls within the framework described here. In
particular, our construction shows that N00N state interferometry is
indeed equivalent to multipass interferometry, as was pointed out
previously \cite{pryde,luis}. The same trick of considering the
(anti-symmetric) Fock space can also be applied to the Fermionic case,
although the Pauli exclusion principle makes Fermionic probes
unattractive. [One may wonder whether the conventional theory of
distinguishable probes is ever applicable, since clearly all probes
will be ultimately be composed of indistinguishable particles.  We
remind that indistinguishable particles are subject to specific
symmetries only regarding their global state. Typically one uses only
certain degrees of freedom of the probe system for measurement
purposes and the remaining degrees of freedom (spin, position, etc.)
can be used to enforce the required symmetry.]

(d)~Extension of quantum metrology: Conventionally, in quantum
metrology the transformation that acts on the probes takes the form
$U_\varphi=e^{i\varphi H}$. Our construction makes it easy to
generalize this to a more general $U'_\varphi=We^{i\varphi H}V$, where
$W$ and $V$ are unitaries independent of $\varphi$. In this case the
sequential strategy cannot be a simple iteration of $U'_\varphi$, as
that will not typically increase the parameter $\varphi$ by $N$ times
(e.g., when $W=\openone$, $V=\sigma_x$ we find that
${U'_\varphi}^2\propto\openone$), but $e^{iN\varphi H}=(W^\dag
U'_\varphi V^\dag)^N$. This suggests that a generalized entangled
parallel strategy consists in the evolution $(W^\dag U'_\varphi
V^\dag)^{\otimes N}(|0\>^N+|1\>^N)$, which corresponds to the
sequential strategy $(W^\dag U'_\varphi V^\dag)^N|\pm\>$.

Is our framework general? The only hypothesis made so far is that
there exists eigenvectors $|0\>$ and $|1\>$ for $H$ connected to the
maximum and minimum eigenvalues. This requirement, certainly true in
finite-dimensional Hilbert spaces, implies that the states $|+\>$ and
$|-\>$ (for the sequential strategy) and the GHZ state $|0\>^N+|1\>^N$
(for the parallel strategy) are ones with maximum spread $\Delta H$,
that saturate the quantum Cramer-Rao bound \cite{qmetr,caves1,caves2}.
Our framework is thus general for finite-dimensional systems, and
allows one to construct new quantum metrology protocols \cite{qmetr}:
all that is needed is a GHZ-type state with a superposition of maximum
and minimum eigenstates of $H$ to achieve a quantum metrology protocol
to estimate the parameter $\varphi$ of which $H$ is the generator of
translations.

(e)~The non-asymptotic case: all results given up to now (and most of
the literature) refer to the asymptotic case where the number of
repetitions $\nu\to\infty$. However, thanks to our construction, it is
possible to adapt the non-asymptotic results
\cite{wiseman,wiseman2,pryde,mitchell} obtained for the sequential
estimation (by adapting the Kitaev phase-estimation algorithms) also
to parallel entangled strategies. This equivalence was already pointed
out in Ref.~\cite{wiseman}: all these strategies employ multiple
rounds of sequential iterations or, equivalently, multiple rounds of
N00N states with different $N$. A completely different strategy is
given in \cite{hayashi}.

(f)~Bandwidth-noise tradeoff: our construction clarifies that the true
power of entanglement in quantum metrology is in allowing one to
achieve the same $\sqrt{N}$ precision enhancement of the sequential
strategy, while keeping the same $N$-fold sampling-time gain of the
parallel strategy over the sequential one \cite{brausmil}: high
precision in short time.

\section{Conclusions}
We presented a simple construction that shows why entanglement is
necessary in quantum metrology: to transform a parallel strategy into
a sequential one, we need perfect correlation in complementary bases:
both in the eigenbasis $|0,1\>$ of the generator $H$ and in its
complementary $|\pm\>$ basis.  Such correlation requires entanglement.
We also presented some applications: our construction easily derives
some new and some known results in quantum metrology.

\section{Acknowledgments}
I acknowledge useful discussions with V. Giovannetti and support by
the Royal Society for attending the International Scientific Seminar
on Sources and signatures of quantum enhancements in technology.

\end{document}